# Spin dynamics of anisotropic azimuthal modes in heterogeneous magnetic nanodisks

**Huirong Zhao[1,2,3], Zhengyun Lei[1,2,3] and Ruifang Wang[1,2,3,4]**

[1] Department of Physics, Xiamen University, Xiamen 361005, China

[2] Institute of Theoretical Physics and Astrophysics, Xiamen University, Xiamen 361005, China

[3] Collaborative Innovation Centre for Optoelectronic Semiconductors and Efficient Devices, Xiamen University, Xiamen 361005, China

4. Key laboratory of low dimensional condensed matter physics, Department of Education of Fujian Province, Xiamen University, Xiamen 361005, China

Email: wangrf@xmu.edu.cn

## Abstract

It is well known that azimuthal spin wave modes of magnetic vortex state in permalloy nanodisks have circular symmetry. Intuitively, magnetic materials having magnetocrystalline anisotropy is not compatible with the circular symmetry of the azimuthal modes. In this article, however, we report cubic azimuthal modes in heterogeneous nanodisks consisting of a permalloy core and a Fe shell. The fourfold symmetry of azimuthal modes is due to the exchange, and magneto-static, interactions between the permalloy core and the Fe shell. In comparison to results of circular azimuthal mode, the vortex switching occurs considerably faster under the excitation of cubic azimuthal mode. The gyration path of vortex core turns into square under the influence of induced cubic anisotropy in the Py region. We find out periodic oscillation of the vortex core size and the gyration speed as well. Our findings may offer a new route for spintronic applications using heterogeneous magnetic nanostructures.

# 1. Introduction

Magnetic vortex (MV) is a common ground state for soft ferromagnetic thin-film structures, ranging from a few microns down to 100 nanometers or so.[1,2] In a MV, the magnetization circulates in-plane around a core where the magnetization orients perpendicular to the sample plane. The vortex core (VC) magnetization can be either upward or downward, denoting the vortex polarity $p = +1$ or $p = -1$, respectively. Spin dynamics of VC reversal under the excitation of ac magnetic field[3-8] and spin current[9-11] receives intensive research interest recently, because of its importance in fundamental physics and technological applications.[1,12] Spin wave excitation and its interaction with the VC play a key role in the dynamic reversal of VC polarity. A MV state features three kinds of spin wave modes, namely the gyrotropic[13-15], azimuthal[16-18] and radial[16,17,19,20]. The sub-GHz gyrotropic mode is characterized by the VC's spiral motion under the excitation of an in-plane ac field. The azimuthal and radial modes, on the other hand, have much higher eigenfrequencies (~10 GHz). Thus, the azimuthal and radial modes can drive VC switching ten times faster than the gyrotropic one.[5,18]

The azimuthal spin wave modes are characterized by their radial ($n = 1, 2, 3, …$) and azimuthal mode numbers ($m = ±1, ±2, ±3, ...$). The clockwise (CW) and counterclockwise (CCW) azimuthal modes correspond to $m < 0$ and $m > 0$, respectively. According to conventional knowledge, only materials with zero magnetocrystalline anisotropy are suitable for the azimuthal spin wave, since the existence of anisotropic energy will suppress the transmission of this mode along azimuthal direction. Up to now, studies on azimuthal spin wave have been focused on permalloy (Py: $Fe_{0.2}Ni_{0.8}$) nanodisks, which have no magnetocrystalline anisotropy.[5,18,21,22] Here, we report novel spin wave dynamics of azimuthal modes in a heterogeneous nanodisk consisting of a Py core and a Fe shell. In contrast to the results in uniform Py disks, our micromagnetic simulations manifest clear fourfold symmetry for the azimuthal spin wave in the Py region, due to the exchange and magneto-static interactions between the Fe shell and Py core. Furthermore, the anisotropic azimuthal spin wave drives considerably faster VC switching, in comparison to the results of homogeneous disks. Meanwhile, the size of VC and the gyration speed oscillate periodically as the VC gyrates along a square path.

## 2. Micromagnetic simulation model

The core-shell nanodisk in study is illustrated in Fig. 1(a). The Py core is 100 nm in radius and the surrounding Fe shell has width of 50 nm. The disk thickness is set as 20 nm. We use typical Py/Fe material parameters for micromagnetic simulations[23] in this paper: saturation magnetization $M_s$ = 800/1714 KA/m, exchange stiffness constant $A_{ex}$ = 13/21 pJ/m, magnetocrystalline anisotropy $K_c$ = 0/47 KJ/m³ and Gilbert damping constant = 0.01/0.01. The exchange stiffness constant across the interface of Py/Fe is set as $A_{ex}^{int}$ = 16 pJ/m, which is the harmonic mean of $A_{ex}$ of Fe and Py. Experimentally, the core-shell nanodisks can be prepared using advanced nanofabrication techniques.[24] The mesh cell size for the simulations is 1.5 × 1.5 × 20 nm³. Additional simulations using cell size of 1.5 × 1.5 × 10 nm³ verified the reliability of results obtained in this paper.

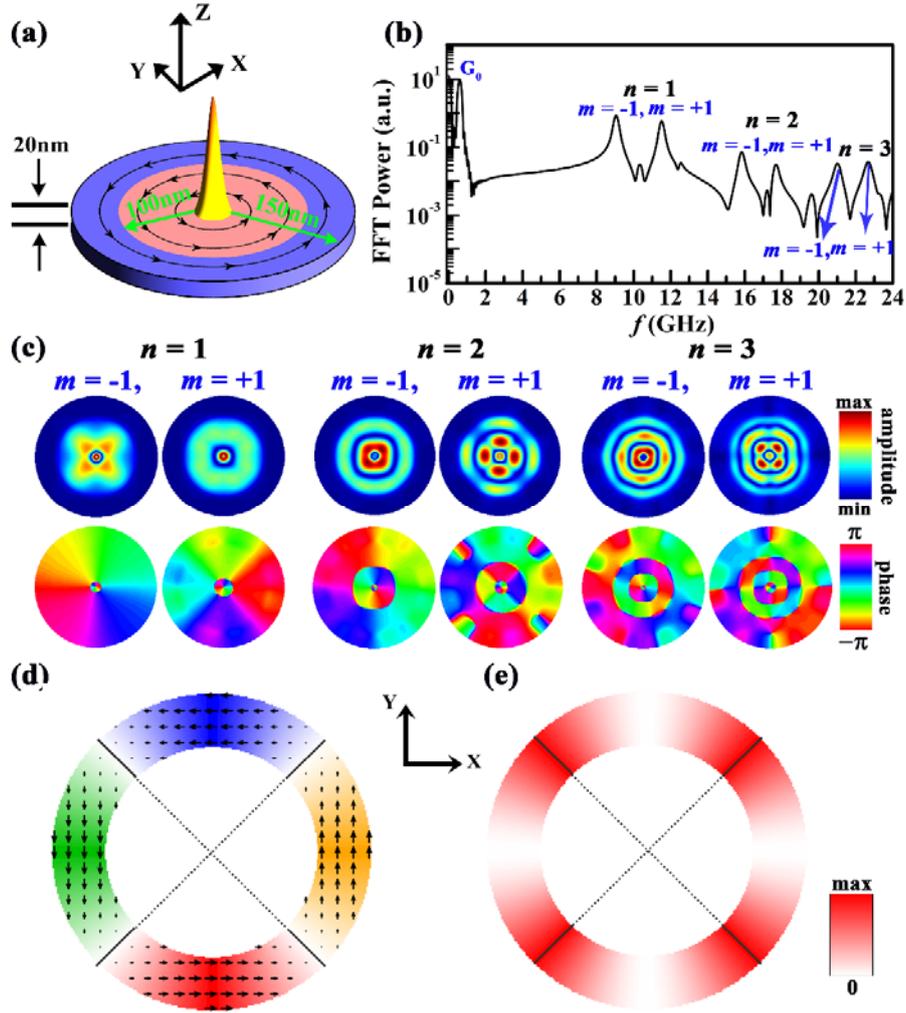

Fig. 1 (a) Schematic view of the core-shell nanodisk with indicated dimensions. The Py core (in magenta) has a radius of 100 nm and the Fe shell (in blue) is 50 nm in width. In its initial state, the magnetic vortex has counterclockwise magnetization rotation and VC polarity $p = +1$. (b) FFT power spectrum of the damped oscillation of $<m_x>$ after the excitation of an in-plane sinc-function field. (c) FFT amplitude (top row) and phase (bottom row) images of the six azimuthal modes. (d) - (e) Distribution of the in-plane components of the magnetocrystalline anisotropy field (d) and the anisotropy energy density (e). In (d), arrows and color shades denotes the amplitude and directions of the magnetocrystalline anisotropy field, respectively. In (e), the color shade indicates the magnitude of the anisotropy energy density. The solid black lines correspond to locations with the lowest magnetocrystalline anisotropy field in (d), and the highest anisotropy energy density in (e). The cubic symmetry of both the anisotropy field and anisotropy energy density of the Fe shell imprints into the Py core, as illustrated by the dotted black lines that connect with the solid black lines.

## 3. Azimuthal modes with fourfold symmetry

The sample is initially in a relaxed ground state with CCW magnetization curling sense and VC polarity $p = +1$, as shown in Fig. 1(a). The diameter of the VC is 11.9 nm, which is defined in such a way that the z component of magnetization ($m_z = M_z / M_s$) at the edge of core is 30% lower than that at the peak. To stimulate the spin wave modes of the vortex state, we apply an in-plane sinc-function field $\vec{B}(t) = \mathbf{e}_x B_0 \sin[2\pi f(t-t_0)]/[2\pi f(t-t_0)]$ with $B_0 = 1$ mT, $t_0 = 1$ ns and $f = 50$ GHz. We then conduct fast Fourier transformation (FFT) on the subsequent temporal oscillation of $<m_x> = <M_x>/M_s$, namely the x-component of magnetization averaged over the entire disk. The resultant FFT power spectrum shown in Fig. 1(b) contains seven resonance peaks. The peak denoted as $G_0$ at 0.595 GHz corresponds to the gyrotropic mode. The other peaks are found at 9.0, 11.5, 15.8, 17.7, 21.0, and 22.7 GHz, respectively, which correspond to the eigenfrequencies of $(n, m) = (1, -1)$, $(1, +1)$, $(2, -1)$, $(2, +1)$, $(3, -1)$, and $(3, +1)$ azimuthal modes, respectively. The small peaks in Fig. 1(b) originate from irregular standing waves, which will disappear if we set $k_c = 0$ for the Fe shell.

Figure 1(c) displays the FFT amplitude and phase images of the azimuthal modes. To understand the fourfold symmetry shown in Fig. 1(c), we plot in Figs. 1(d) and 1(e) the spatial distribution of the magnetocrystalline anisotropy field and the anisotropy energy density[25] respectively, for the Fe shell in the relaxed ground state. The cubic anisotropy field in the Fe shell is expressed as $B_{ani} = (2k_c/M_s)\cos\theta\cos 2\theta$, in which $\theta$ is the angle between magnetization and the nearest easy axis, which is defined as the x, y and z axes in the sample. The maximum anisotropy field, i.e., $2k_c/M_s = 54.8\ mT$, is found along the easy axes. Whereas the minimum, $B_{ani} = 0$, occurs at $\theta = \pm\pi/4$ or $\pm 3\pi/4$. The density of anisotropy energy is defined as $E_{ani} = k_c(\alpha_1^2\alpha_2^2 + \alpha_2^2\alpha_3^2 + \alpha_3^2\alpha_1^2)$, where $\alpha_1$, $\alpha_2$, $\alpha_3$ are the direction cosines of magnetization with respect to the easy axes. This directional dependence of the crystalline anisotropy energy is disadvantageous for transmission of azimuthal spin waves in the Fe shell, because of the energy barriers at $\theta = \pm\pi/4$ and $\pm 3\pi/4$. As a result, the Py core has much larger FFT amplitude, i.e. stronger spin wave oscillation, than the Fe shell, as shown in Fig. 1(c). On the other hand, the azimuthal modes in the Py core shows clear fourfold symmetry in Fig. 1(c). This result indicates that the Fe shell induces cubic anisotropy in the Py core, due to the exchange and magneto-static coupling between the core and the shell. This induced cubic symmetry results in nontrivial spin dynamics that we will discuss in detail below.

## 4. Dynamics of the anisotropic azimuthal mode

After identifying the azimuthal modes in this core-shell nanodisk, we apply an in-plane CW rotating field with amplitude of 8 mT and frequency of 9 GHz to excite the ($n = 1$, $m = -1$) mode. The snapshots of the $m_z$ spatial distribution shown in Fig. 2 illustrates the spin dynamics until the VC switching occurs. At $t = 0$ ps, the vortex is in the relaxed ground state. After excitation by the in-plane rotating field, the VC moves away from the disk center due to the interaction of azimuthal spin wave with the gyrotropic mode.[5,22] For the $m = -1$ mode in study, the gyrofield[26] and spin wave oscillation acts constructively, thus forms relatively broad negative and positive $m_z$ regions, to the left and right side of the core, respectively, in relation to the direction of motion.[5] Note that the VC, along with the negative and positive $m_z$ regions, deforms periodically whenever they travel across the anisotropic energy barriers. Specifically, they are compressed (or stretched) when they are approaching (or leaving) the barriers at $\theta = \pm \pi/4$ and $\pm 3\pi/4$, in order for the system to keep its total energy low. A vortex-antivortex pair[8,26] is created at $t = 689$ ps, inside the negative $m_z$ region as the spin wave oscillation intensifies. The antivortex then annihilates with the original vortex core at $t = 751$ ps, leaving the surviving vortex with reverse polarity in the dot. This completes the VC reversal process. In comparison, VC reversal in a homogeneous Py disk of same dimensions takes 1056 ps, driven by a 9 mT resonant field that excites the ($n = 1$, $m = -1$) mode.[18]

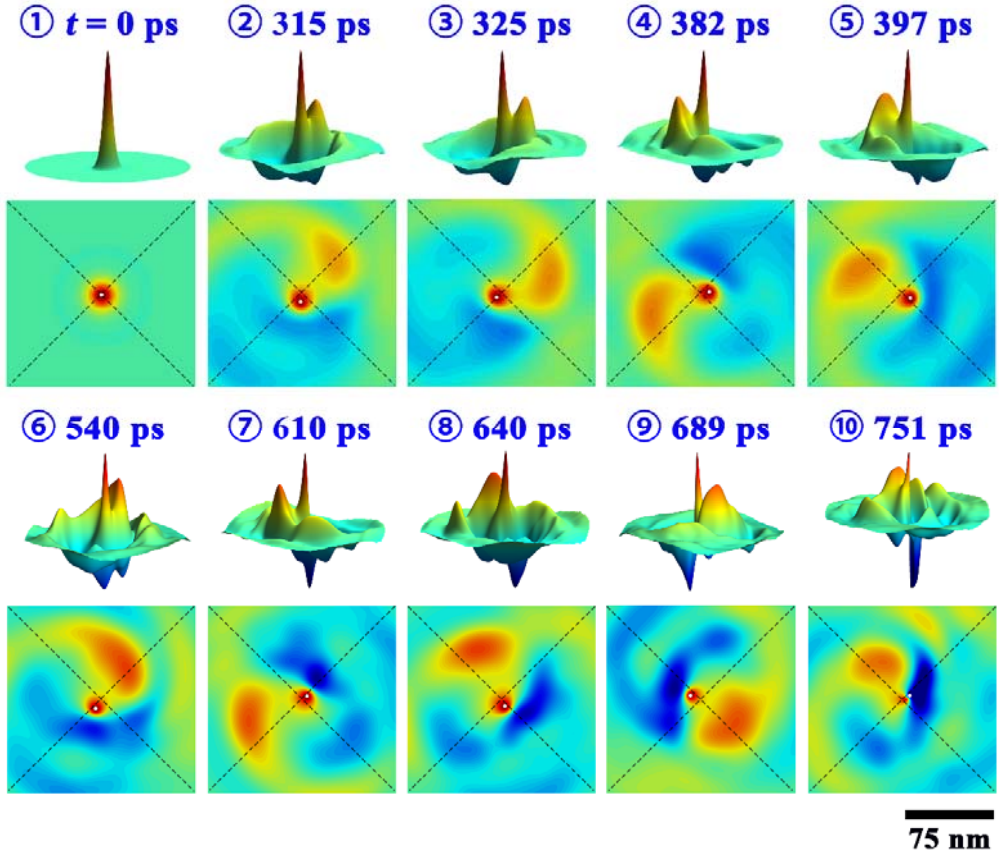

Fig. 2 Snapshot images showing spin dynamics of the magnetic vortex after application of a resonant in-plane rotating field to excite the ($n = 1$, $m = -1$) mode. Below each snapshot is top view of a $150 \times 150$ nm$^2$ area in the center of the disk. The apex of VC is marked as a white dot. The dotted diagonal lines indicate induced anisotropic energy barriers in the Py core.

The gyration trajectory of VC, illustrated in Fig. 3(a), manifests an unusual feature of the anisotropic azimuthal mode. We plot the trajectory by recording positions of the apex of the VC. It shows that the VC spirals around the disk center in a clockwise sense at a frequency of 8.8 GHz, slightly less than the eigenfrequency of the $m = -1$ mode. This behavior is consistent with previous studies on homogeneous vortices.[18] However, in contrast to the circular gyration in homogeneous disks[5,18], the VC trajectory in Fig. 3(a) transforms from circular into square after $t = 294$ ps, when the VC is about 6.4 nm away from the disk center. Such transformation indicates that the VC interacts with growing induced cubic anisotropy as it moves toward the Py/Fe interface.

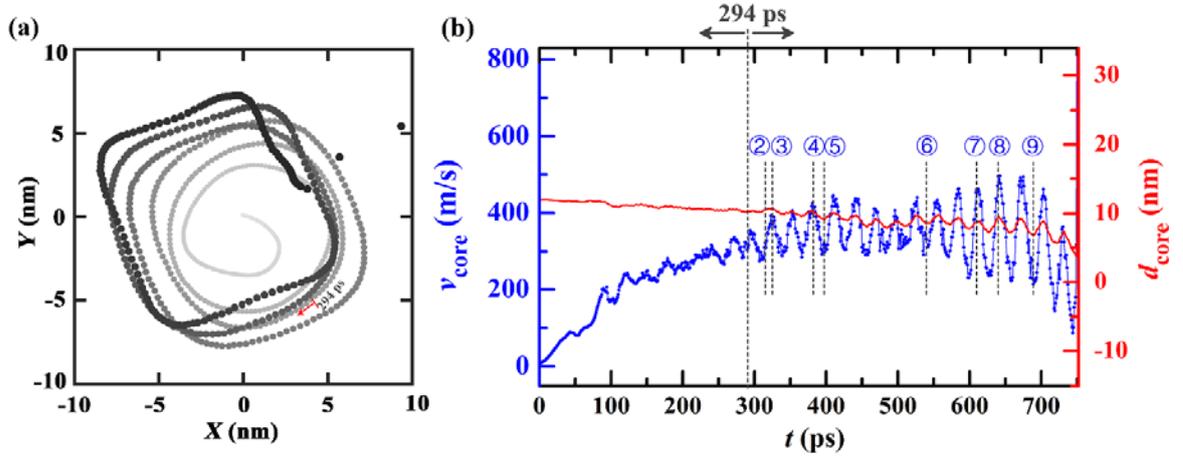

Fig. 3 (a) The VC trajectory under the excitation of the ($n = 1$, $m = -1$) mode. (b) Variations of gyration speed and the size of VC with time. The circled numbers in (b) correspond to the snapshot images with same numbering in figure 2.

Furthermore, we show temporal evolution of the gyration speed ($v_{core}$) and diameter ($d_{core}$) of VC in Fig. 3(b). Before $t = 294$ ps, $v_{core}$ increases in general with time, while showing some irregular oscillations. At the same time, $d_{core}$ decreases almost monotonically. These behaviors are similar to the results obtained in homogeneous nanodisks[5,18,21], since the induced cubic anisotropy is weak in close proximity to the disk center. At $t = 294$ ps, the core velocity has increased to 351 m/s and the core diameter is reduced to 10.3 nm. After $t = 294$ ps, however, both $v_{core}$ and $d_{core}$ oscillates with frequency of 34.4 GHz, which is exactly four times as large the gyration frequency of VC. The reason is that induced cubic anisotropy in Py causes VC deceleration and compression, when the VC is approaching energy barriers at $\theta = \pm \pi/4$ and $\pm 3\pi/4$; the opposite is true when the VC is leaving the barriers.

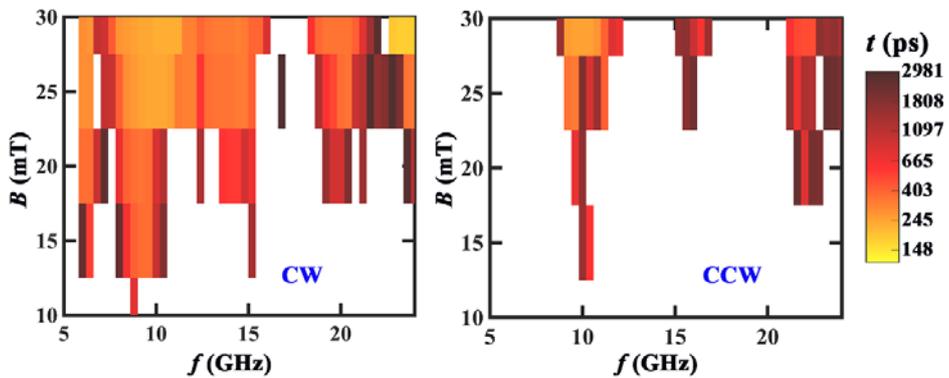

Fig. 4 Phase diagrams of VC reversal under the excitation of clockwise (left) and counterclockwise (right) rotating fields, respectively. The logarithmic color scale denotes excitation time until VC switching.

Finally, Fig. 4 shows how long for the VC to reverse after a rotating field is applied. The field amplitude varies from 10 mT to 30 mT, and the field frequency keeps in the range of 6 to 24 GHz. As expected, the VC switching occurs faster for higher field amplitude. The local minima coincide with the eigenfrequencies of azimuthal modes, a clear indication of resonant behavior. Note that the gyrofield acts constructively with the spin wave for the $m = -1$ mode, but destructively for $m = +1$.[5] Therefore, VC switching is faster and occurs in wider frequency range under excitation of the $m = -1$ mode. However, for field amplitude above 25 mT, the ($n = 1$, $m = +1$) mode dominates and the unidirectional reversal of VC is lost.[5]

In summary, azimuthal modes in heterogeneous magnetic nanodisks show clear fourfold symmetry, due to interactions between the Fe shell and the Py core. The large difference in crystalline anisotropy across the Py/Fe interface causes confinement of azimuthal spin wave inside the Py region. We observe markedly faster VC switching, in comparison to the results of homogeneous disks. The VC gyration trajectory turns into square under the influence of induced cubic anisotropy in Py, which results in periodic oscillations of both the gyration speed and size of the VC as well.

We are grateful for the financial support from the National Natural Science Foundation of China under Grant Nos. 10974163 and 11174238.